\documentclass[11pt]{article} \usepackage{hyperref} \pdfoutput=1
\begin{document}

\title{Transport and mixing in thin films of oxytactic bacteria}
\author{Barath Ezhilan, Amir Alizadeh Pahlavan \& David Saintillan \\ 
 \\
 \vspace{6pt} Department of Mechanical Science and Engineering, \\ 
University of Illinois at Urbana-Champaign, Urbana, IL 61801, USA}
\maketitle
\begin{abstract} 
 
This fluid dynamics  video presents three-dimensional kinetic simulations of the dynamics in suspensions of oxytactic bacteria confined in thin liquid film surrounded by air. At the initial time, the bacterial concentration is uniform and isotropic,  and there is no oxygen inside the film. The spatio-temporal dynamics of the oxygen and bacterial concentration are analyzed. For small film thicknesses, there is a weak migration of bacteria to the boundaries, and the oxygen concentration is high inside the film as a result of diffusion; both bacterial and oxygen concentrations quickly reach steady states. Above a critical film thickness, a transition to chaotic dynamics is observed and is characterized by turbulent-like 3D motion, the formation of bacterial plumes, enhanced oxygen mixing and transport into the film, and hydrodynamic velocities of magnitude up to 7 times the single bacterial swimming speed. This collective motion arises as a result of the combined effect of hydrodynamic interactions and oxygentaxis, and may play a role in facilitating the access of bacteria to oxygen in fluid environments.
 
 \end{abstract}
 
  \end{document}